\documentclass[aps,psfig,epsfig,preprint] {revtex4}
\usepackage{epsfig}
\usepackage{color}
\usepackage{bm}
\usepackage{subfigure}

\begin{document}
\draft
\title{
{\normalsize \hskip4.2in USTC-ICTS-07-04} \\{\bf Y(2175):
Distinguish Hybrid State from Higher Quarkonium}}

\author{Gui-Jun Ding\footnote{E-mail: dinggj@mail.ustc.edu.cn}, Mu-Lin
Yan\footnote{E-mail: mlyan@ustc.edu.cn}}

\affiliation{\centerline{Interdisciplinary Center for Theoretical
Study and Department of Modern Physics,} \centerline{University of
Science and Technology of China,Hefei, Anhui 230026, China} }

\begin{abstract}
The possibility of $Y(2175)$ as a $2{\;^3D_1}$ $s\bar{s}$ meson is
studied. We study the decay of $2{\;^3D_1}$ $s\bar{s}$ from both the
$^3P_0$ model and the flux tube model, and the results are similar
in the two models. We show that the decay patterns of $1^{--}$
strangeonium hybrid and $2{\;^3D_1}$ $s\bar{s}$ are very different.
The experimental search of the decay modes $KK$, $K^{*}K^{*}$,
$K(1460)K$, $h_1(1380)\eta$ is suggested to distinguish the two
pictures. Measuring the $K^{*}K^{*}$ partial width ratios is crucial
to discriminate the $2{\;^3D_1}$ from the $3{\;^3S_1}$ $s\bar{s}$
assignment.

\vskip 0.5cm

PACS numbers: 12.39.Mk, 13.20.Gd, 13.25.-k, 14.40.Gx
\end{abstract}
\maketitle
\section{introduction}
Recently the Babar Collaboration has observed a structure at 2175
MeV in $e^{+}e^{-}\rightarrow\phi f_0(980)$ via initial-state
radiation, which is consistent with $1^{--}$ resonance with a mass
$m=2.175\pm0.010\pm0.015 \rm{GeV/c^2}$ and width
$\Gamma=58\pm16\pm20 \rm{MeV}$\cite{babar}. Henceforth, this
structure is denoted as $Y(2175)$. Furthermore, the Babar
collaboration demonstrates that this structure is not due to the
dominant $K^{*}(892)K\pi$ intermediate states, and there is no known
meson resonance with $I=0$ near this mass.

In Ref.\cite{ding}, we suggested that this structure is a $1^{--}$
strangeonium hybrid($s\bar{s}g$) and the decay properties are
studied from both the flux tube model and the constitute gluon
model. Both the mass and decay width are consistent with the current
experimental data in the hybrid scenario. Moreover, we suggested
that the tetraquark hypothesis for Y(2175)\cite{wang} is not favored
by the current data in\cite{babar}, although this picture cannot be
completely excluded. The mass of $Y$(2175) which is assigned as a
tetraquark state, has been calculated in QCD sum rule\cite{wang}. To
confirm that $Y(2175)$ is hybrid or another exotic state, it is
necessary to examine the radial excitation of quarkonium in the
quark model to determine whether they imitate the decay and
production properties of the hybrid state. The aim of this work is
to study the decay of $1^{--}$ strange quarkonium(2$\;^3D_1$) and
reveal possible experimental signals that can discriminate between
the hybrid and the higher quarkonium.

Its quantum numbers $J^{PC}=1^{--}$ implies that the possible
quarkonium states are
$^3S_1,\;^3D_1,\;2\;^3S_1,\;2\;^3D_1,\;3\;^3S_1,\;3\;^3D_1$ (in the
notation $n\;^{2S+1}L_J$, denoting the $nth$ state with spin $S$,
orbital angular momentum $L$, and total angular momentum $J$) and so
on. Among these states, only the $3\;^3S_1$ and $2\;^3D_1$ strange
quarkonium states have masses consistent with $Y(2175)$ within the
experimental error\cite{isgur1}. The decay of $3\;^3S_1$ strange
quarkonium has been studied in detail in the $^3P_0$ model by
T.Barnes et al.,\cite{barnes1,barnes2}, and this state is predicted
to be a rather broad resonance, $\Gamma\approx380$MeV in the $^3P_0$
model, and it mainly decays into
$K^{*}K^{*},\;KK^{*}(1414),\;KK_1(1273)$. However, the $1^{--}$
strangeonium hybrid that we predicted is much narrower
$\Gamma\approx100-130$MeV, and the $K^{*}K^{*}$ mode is
forbidden\cite{ding}. Since the width of $3\;^3S_1$ strange
quarkonium is much larger than that of
$Y(2175)$($\Gamma=58\pm16\pm20 \rm{MeV}$), it is very difficult to
identify $Y(2175)$ as a $3\;^3S_1$ strange quarkonium.

In this paper, we will calculate the partial widths of $2\;^3D_1$
strange quarkonium to all Okubo-Zweig-Iizuka-(OZI) allowed two- body
final states allowed by phase space. To assess the correctness of
our analysis, we calculated the widths by using both the $^3P_0$
decay model\cite{micu,orsay,roberts,godfrey,barnes3} and the flux
tube breaking model\cite{isgur2,swanson,page}. Our goal is to shed
some light on the nature of $Y(2175)$ and reveal some promising
signals that can discriminate the $1^{--}$ strangeonium hybrid
picture form the $2\;^3D_1$ strange quarkonium.

In this paper, we will study the decay of $2\;^3D_1$ strange
quarkonium from the $^3P_0$ model in Sec. II and from the flux tube
model in Sec. III. We use the simple harmonic approximation in both
models so that the amplitudes can be derived analytically. Finally,
we present our summary and some discussions.

\section{The decay of $2\;^3D_1$ $s\bar{s}$ in $^3P_0$ decay model}

The $^3P_0$ model (quark pair creation model), which describes the
process that a pair of quark-antiquark with quantum number $J^{PC} =
0^{++}$ is created from the vacuum was first proposed by
Micu\cite{micu} in 1969. In the 1970s, this model was developed by
Yaouanc et al. \cite{orsay} and applied to study hadron decays
extensively. The $^3P_0$ model is applicable to OZI-allowed strong
decay of a meson into two other mesons, as well as the two
body-decay of baryons and other hadrons\cite{isgur3}. In the $^3P_0$
model, the created quark pairs with any color and any flavor can be
generated anywhere in space, but only those whose color-flavor wave
functions and spatial wave functions overlap with those of outgoing
hadrons can make a contribution to the final decay width.

It is widely assumed that $^3P_0$ model is successful because it
provides a good description of many of the observed decay amplitudes
and partial widths of open flavor meson strong decays. Several
published papers study the decay of light mesons, open charm mesons
and charmonium using different wavefunction and phase space
normalization\cite{qpc1,qpc2}. We will use the diagrammatic
technique developed in Ref.\cite{barnes3} to derive the amplitudes
and the $^3P_0$ matrix elements. In this formalism, the $^3P_0$
model describes decay matrix elements using a $q\bar{q}$ pair
production Hamiltonian, which is the nonrelativistic limit of,
\begin{equation}
\label{1}H_I=g\int
d^3\mathbf{x}\;\overline{\psi}(\mathbf{x})\;\psi(\mathbf{x})
\end{equation}
where $\psi$ is a Dirac quark field, $g=2m_q\gamma$, $\gamma$ is the
strength of the conventional $^3P_0$ mode, and $m_q$ is the mass of
both created quarks. To determine a decay rate, we evaluate the
matrix element of the decay Hamiltonian between the initial and
final states,
\begin{equation}
\label{2}\langle
BC|H_I|A\rangle=h_{fi}\;\delta^3(\mathbf{p}_{A}-\mathbf{p}_{B}-\mathbf{p}_{C})
\end{equation}
To compare with the experiment, we transform the helicity $h_{fi}$
into the partial wave amplitude ${\cal{M}}_{L_{BC},S_{BC}}$ by the
recoupling calculation\cite{jacob} . Then the decay width is:
\begin{equation}
\label{3}\Gamma(A\rightarrow
B+C)=2\pi\frac{PE_BE_C}{M_A}\sum_{L,S}|{\cal{M}}_{LS}|^2
\end{equation}
where relativistic phase space normalization has been taken, and $P$
is the momentum of the final states in the rest frame of $A$, i.e.,
$P=\frac{[(M^2_A-(M_B+M_C)^2)(M^2_A-(M_B-M_C)^2)]^{1/2}}{2M_A}$. The
decay amplitudes for $2\;{^3D_1}\rightarrow {^1S_0}+{^1S_0}$,
$2\;{^3D_1}\rightarrow {^3S_1}+{^1S_0}$ etc can be derived
analytically  under the simple harmonic oscillator wavefunction
approximation. The mass difference $m_{u,\;d}\neq m_s$ is ignored in
the wavefunctions as Ref.\cite{barnes1,barnes2}. However, this
approximation may not be good in some channels, and we will clearly
see the effect of large strange quark mass in the flux tube model in
Sec.III. In Ref.\cite{barnes1,barnes2}, the authors assume that the
harmonic oscillation parameter $\beta_A$ of the initial state is the
same as the $\beta$ of the final states. In the present paper, we
relax this assumption by allowing $\beta_A$ to be different from
$\beta$.

We assume that the harmonic oscillation parameter of the final
states and the pair-production amplitude respectively are $\beta=0.4
\rm{GeV},\; \gamma=0.4$ as in Ref.\cite{barnes1,barnes2}. It is
theoretically expected that the magnitude of $\beta_A$ for higher
excited states are smaller than those of lower states by $\leq 0.1$
GeV, so we take $\beta_A=0.35$GeV in our numerical calculations of
the partial decay widths. Meson mass are taken from the Particle
Data Group(PDG)-2006\cite{pdg}. If a state was not included in the
Meson Summary Table of PDG, we use an estimated mass motivated by
the spectroscopy predictions\cite{isgur1}, adjusted in the absolute
value relative to the known masses. For the pseudoscalar $\eta$ and
$\eta^{\prime}$, we assume perfect mixing, and the flavor structures
are as follows,
\begin{eqnarray}
\nonumber&&\eta=\frac{1}{2}(u\bar{u}+d\bar{d})-\frac{1}{\sqrt{2}}\;s\bar{s}\\
\label{4}&&\eta^{\prime}=\frac{1}{2}(u\bar{u}+d\bar{d})+\frac{1}{\sqrt{2}}\;s\bar{s}
\end{eqnarray}
The mixing angle is consistent with the angle obtained from the
$\eta-\eta^{\prime}$ mass matrix. Using the analytical decay
amplitudes and including the flavor factors, summing over all final
flavor states, we obtain the numerical value of the partial decay
width for $2\,{^3D_1}$ $s\bar{s}$ decay in the $^3P_0$ model, which
are listed in the second column of the Table I.

\begin{table}[hptb]
\begin{center}
\caption{The decay of Y(2175) as $2\,{^3D_1}$ strange quarkonium in
the $^3P_0$ model and the flux tube model. $\Gamma_{LJ}({\rm MeV})$
is partial decay width, where $L$ represents the relative angular
momentum between two final states, and $J$ is the total angular
momentum of the final states. We choose $\beta=0.4$GeV,
$\beta_A=0.35$GeV. The starred amplitudes vanish exactly with the
simple harmonic oscillator wavefunction. The large difference in
some channels is due to the large strange quark mass of the nodal
suppression effect. For comparing the above predictions for Y(2175)
as a $2\,{^3D_1}$ $s\bar{s}$ state with those of the possible hybrid
interpretation in \cite{ding} and of the possible $3\,^3S_1$
quarkonium assignment in \cite{barnes2}, the corresponding results
of \cite{ding} and \cite{barnes2} are listed in the final two
columns of the table. Note, the mass of $3\;^3S_1$ $s\bar{s}$ state
in \cite{barnes2} was set to be 2050MeV.}
\begin{tabular}{|c|c|c|c|c|}\hline\hline
\multicolumn{3}{|c|}{Y(2175) as $2\;^3D_1\; s\bar{s}$ quarkonium }&
Y(2175) as
 & Y(2175) as $3\;^3S_1$ \\
\multicolumn{3}{|c|}{} & $s\bar{s}\;g$ hybrid \cite{ding} &
$s\bar{s}$ quarkonium \cite{barnes2} \\
\hline Decay modes &   $\Gamma_{LJ}$ in $^3P_0$ Model  &
$\Gamma_{LJ}$ in Flux Tube
Model & in Flux Tube Model & in $^3P_0$ Model\\
\hline  $ KK $   & $\Gamma_{P\;0}=$ 9.8 & $\Gamma_{P\;0}$= 23.1
 & $ 0$ & $  0$ \\ \hline
$ K^{*}K$   & $\Gamma_{P\;1}$= 1.3 & $\Gamma_{P\;0}$= 11.7 & 3.7 &
20
\\\hline

$ \phi\eta$   & $\Gamma_{P\;1}$= 0   & $\Gamma_{P\;1}$= 0 & 1.2
 & 21  \\\hline

$ \phi\eta^{\prime}$ &$\Gamma_{P\;1}$= 2.9 & $\Gamma_{P\;1}$= 2.8 &
0.4 & 11
\\\hline

$ K^{*}K^{*}$       &  $\Gamma_{P\;0}$= 0.76   & $\Gamma_{P\;0}$= 0 & 0 & 102 \\
                                         &  $\Gamma_{P\;1}$= $0^{*}$
                                         & $\Gamma_{P\;1}$=$0^{*}$ & & \\
                                         &  $\Gamma_{P\;2}$= 0.15   & $\Gamma_{P\;2}$= 0
                                         & &  \\
                                         & $\Gamma_{F\;2}$= 17.2  & $\Gamma_{F\;2}$= 23.5
                                         & & \\\hline

$K(1460)K$         & $\Gamma_{P\;0}$= 58.3 &$\Gamma_{P\;0}$= 50.2 &
0 & 29 \\\hline

$K^{*}(1410)K$     & $\Gamma_{P\;1}$= 31.9 & $\Gamma_{P\;1}$= 26.0 &
23 & 93 \\\hline

$h_1(1380)\eta$    & $\Gamma_{S\;1}$= 3.6 & $\Gamma_{S\;1}$= 3.5 & 0
& 8
\\\hline

$ K_1(1270)K$       & $\Gamma_{S\;1}$= 2.3  &
$\Gamma_{S\;1}$= 20.5 & 35.3 & 58 \\

                                         & $\Gamma_{D\;1}$= 19.6  & $\Gamma_{D\;1}$=
25.9 & & \\\hline

$K_1(1400)K$       & $\Gamma_{S\;1}$= 3.0  &
$\Gamma_{S\;1}$= 0.8 & 70.1 & 26 \\

                                         & $\Gamma_{D\;1}$= 5.6  & $\Gamma_{D\;1}$=
8.6 & &\\\hline

$ K_2(1430)K$       & $\Gamma_{D\;2}$= 10.8  & $\Gamma_{D\;2}$= 15.3
& 15.0 & 9 \\\hline

$\Gamma_{tot}$                           & 167.21 &211.9 &148.7& 378
\\\hline\hline
\end{tabular}
\end{center}
\end{table}

The $2\;{^3D_1}$ $s\bar{s}$ state is predicted to be rather narrow
in the $^3P_0$ model, $\Gamma\approx167.2$MeV, so we cannot exclude
the possibility that $Y(2175)$ could be a $2\;{^3D_1}$ $s\bar{s}$
state. In the case of $\beta_A=0.35$GeV, the dominant decay modes
are:
\begin{eqnarray}
\label{5}&&2\;{^3D_1}\rightarrow
K(1460)K,\;K^{*}(1410)K,\;K_1(1270)K,\;K^{*}K^{*}
\end{eqnarray}
All these lead to the important $KK\pi\pi$ final state. We suggested
in \cite{ding} that if Y(2175) is assigned as a hybrid, the decay
modes of $K(1460)K$ and $K^{*}K^{*}$ are forbidden due to the famous
selection rule\cite{hybrid} and since $K(1460)$ is $2{\;^1S_0}$
state, and $K^{*}$ is $S$-wave state. However, when Y(2175) is as
$2{^3D_1}$ strange quarkonium discussed in this present paper, in
contrast with the strangeonium hybrid picture, the widths of decay
modes $K(1460)K$ and $K^{*}K^{*}$  are significantly larger (please
see Table I). This remarkable feature is a criterion to distinguish
the quarkonium interpretation from the hybrid assignment for
Y(2175). Therefore, experimentally observing the $K(1460)K$ and
$K^{*}K^{*}$ decay modes of $Y(2175)$ is crucial for exploring the
nature of $Y(2175)$. The $K^{*}K$, $\phi\eta$ and $\phi\eta'$ modes
are near the nodes of the decay amplitudes, as a result, the decay
widths corresponding to them are predicted to be rather small.
Furthermore, we can see another interesting property that
$2\;{^3D_1}$ $s\bar{s}$ prefers to decay into $2S+1S$ final
states(Table I). Namely, $K^{*}(1410)K$ has a large branch ratio if
the problematical $K^{*}(1410)$ is a $2{\;^3S_1}$ state.

The $K^{*}K^{*}$ mode is especially interesting and there are four
partial widths, i.e., $\Gamma_{P0}$, $\Gamma_{P1}$, $\Gamma_{P2}$
and $\Gamma_{F2}$. If Y(2175) is a pure $2{\;^3D_1}$ $s\bar{s}$
state, we predict $\Gamma_{P1}=0$, $\Gamma_{F2}$ is the largest, and
the ratio $\Gamma_{P2}/\Gamma_{P0}=1/5$, which is independent of the
radial wavefunction (in Table I $\Gamma_{P2}/\Gamma_{P0}=0.197$,
because we keep only two effective numbers in numerical results of
partial widths there). However, if Y(2175) is assigned as a
$3{\;^3S_1}$ $s\bar{s}$ state, both $\Gamma_{P1}$ and $\Gamma_{F2}$
are predicted to be zero, and $\Gamma_{P2}/\Gamma_{P0}=20$.
Determining the $K^{*}K^{*}$ decay width ratios is very important to
examine whether Y(2175) is a $2{\;^3D_1}$ or $3{\;^3S_1}$ $s\bar{s}$
state.

To test the robustness of the these conclusions, we should study the
stability of these results  with respect to independent variations
in $\beta_A$ and the mass of the initial state. We show the
$\beta_A$ dependence of the partial widths and total width
respectively in Fig.1 and Fig.2. We can see that the width of
$2{\;^3D_1\rightarrow K_2(1430)K}$ depends weakly on $\beta_A$,
however, the partial width of the modes $K(1460)K$, $K^{*}(1410)K$,
$K_1(1400)K$, $K^{*}K^{*}$ and $h_1(1380)\eta$ vary dramatically
with $\beta_A$. For small $\beta_A$($\beta_A\simeq0.3\sim0.35$),
$2\;{^3D_1}$ $s\bar{s}$ dominantly decays into $K(1460)K$,
$K^{*}(1410)K$, $K_1(1270)K$, $K^{*}K^{*}$ and $KK$, while it
dominantly decays into $K_1(1400)K$, $h_1(1380)\eta$, $K^{*}K^{*}$,
$K_1(1270)K$ for large $\beta_A$ ($\beta_A\simeq0.4\sim0.5$).
Experimental observations of $K(1460)K$, $K^{*}K^{*}$ or $KK$ modes
would be strong indications of a $2\;{^3D_1}$ $s\bar{s}$ component.
Since a $1^{--}$ strangeonium hybrid decay into $h_1(1380)\eta$ is
forbidden due to the "spin selection" rule\cite{hybrid}, the
$h_1(1380)\eta$ mode is also very important in determining the
nature of $Y(2175)$.

\begin{figure}[hptb]
\centering
\begin{minipage}[t]{0.46\textwidth}
\centering
\includegraphics[width=8cm]{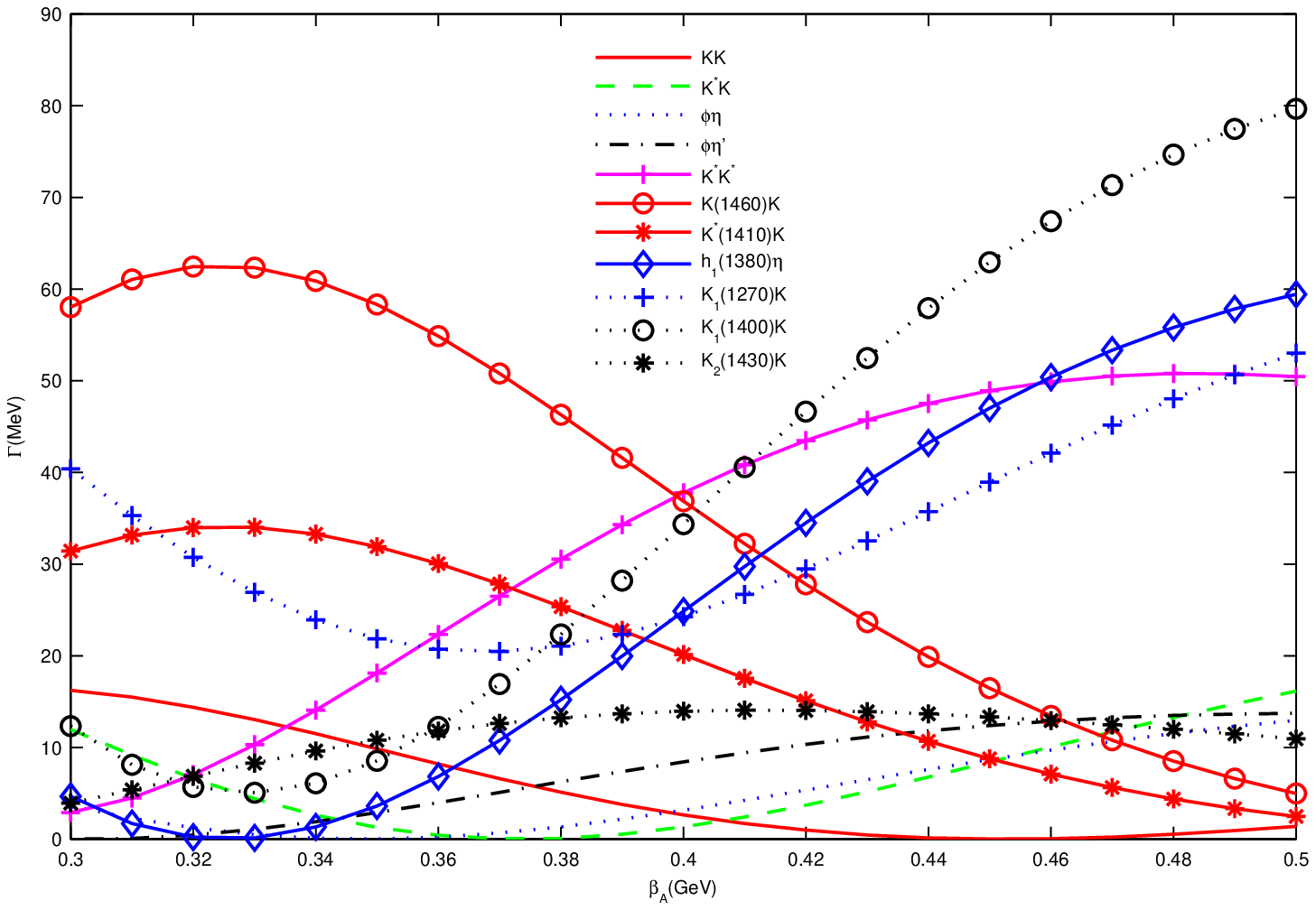}
\caption{The variation of Y(2175) partial decay widths with
$\beta_A$ as a $2{\;^3D_1}$ $s\bar{s}$ state in the $^3P_0$ model.}
\end{minipage}%
\hspace{0.04\textwidth}%
\begin{minipage}[t]{0.46\textwidth}
\centering
\includegraphics[width=8cm]{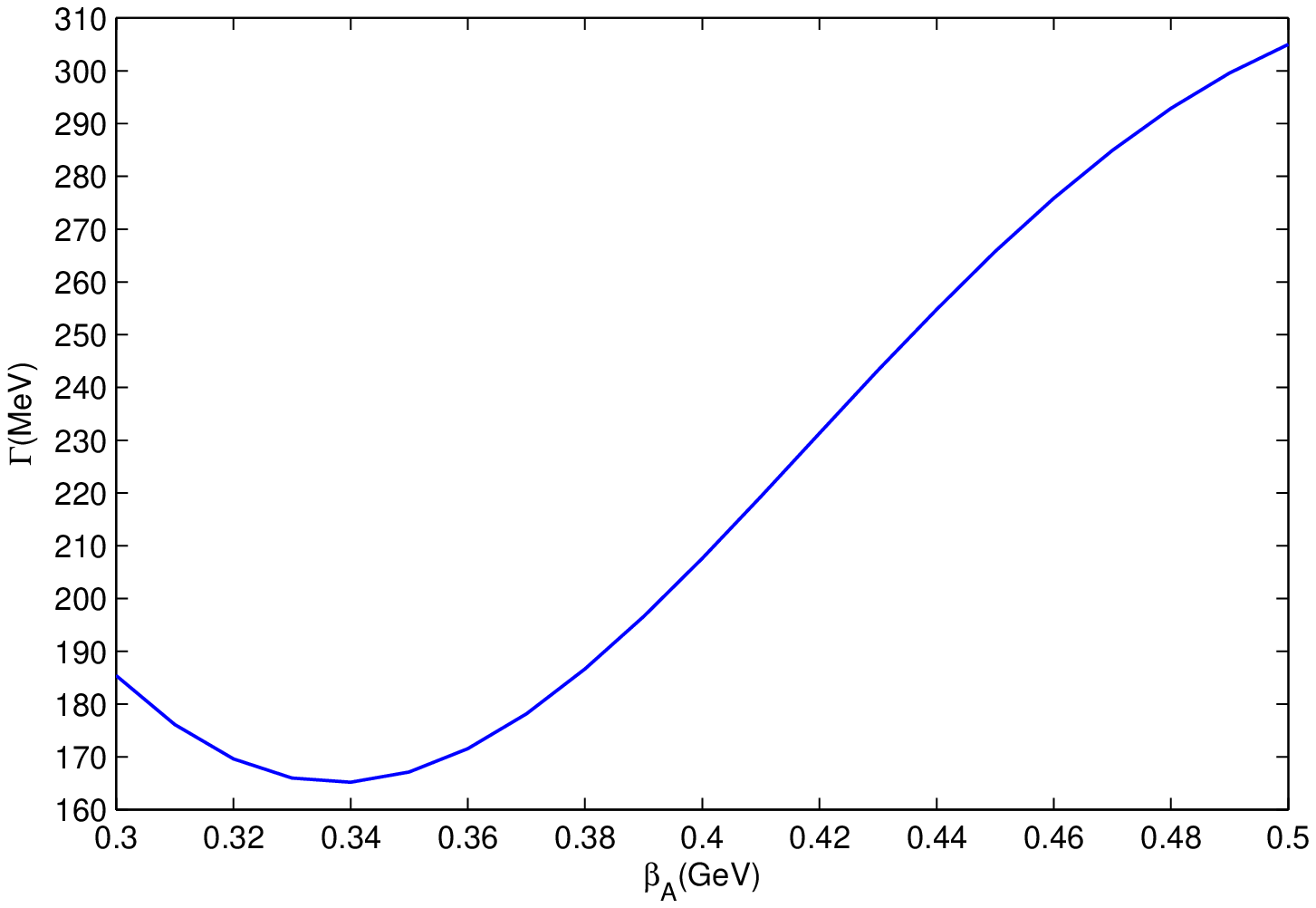}
\caption{Y(2175) total width dependence on $\beta_A$ as a
$2{\;^3D_1}$ $s\bar{s}$ state in the $^3P_0$ model.}
\end{minipage}
\end{figure}

The variation of the partial decay width and total width with the
mass of the initial state are shown in Fig.3 and Fig.4,
respectively, where the initial state mass is denoted as $M_A$, and
we choose $\beta_A=0.35$ GeV. The total width becomes large with
increasing $M_A$. Interestingly, the partial decay widths
$2{\;^3D_1}\rightarrow K_1(1400)K$ and $2{\;^3D_1}\rightarrow
h_1(1380)\eta$ decrease when the mass of $2{\;^3D_1}$ state
increases. This is because the modes $K_1(1400)K$ and
$h_1(1380)\eta$ are closer to the nodes of the decay amplitude with
increasing $M_A$. Moreover, both the $K(1460)K$ and $K^{*}K^{*}$
modes always have a sizable branch ratio in the mass region
$2.05\sim2.25$ GeV.

\begin{figure}[hptb]
\centering
\begin{minipage}[t]{0.46\textwidth}
\centering
\includegraphics[width=8cm]{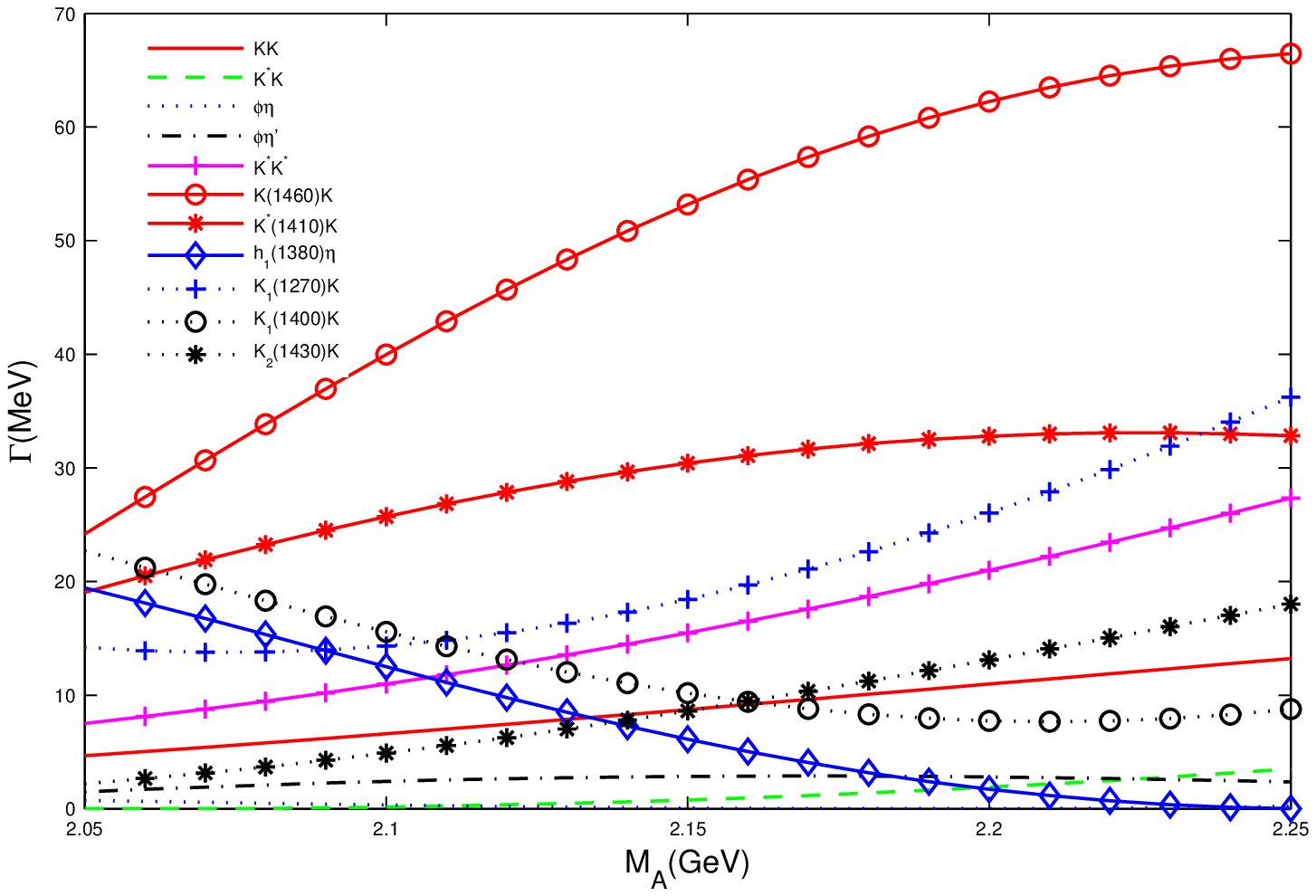}
\caption{The variation of Y(2175) partial decay widths with
$M_A$(the initial state mass) as a $2{\;^3D_1}$ $s\bar{s}$ state in
the $^3P_0$ model.}
\end{minipage}%
\hspace{0.04\textwidth}%
\begin{minipage}[t]{0.46\textwidth}
\centering
\includegraphics[width=8cm]{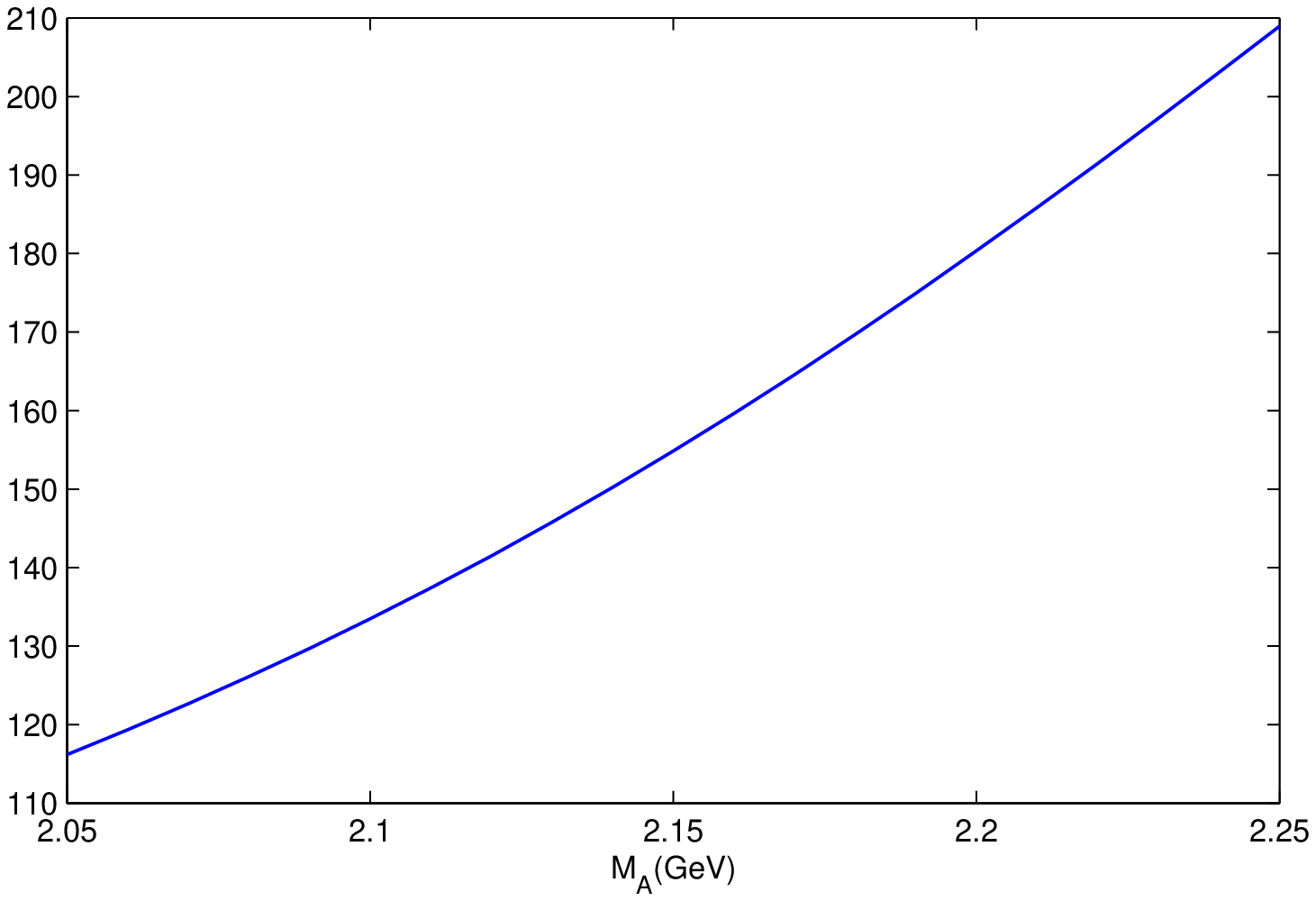}
\caption{Y(2175) total width dependence on $M_A$ as a $2{\;^3D_1}$
$s\bar{s}$ state in the $^3P_0$ model.}
\end{minipage}
\end{figure}

\section{decay of $2\;{^3D_1}$ $s\bar{s}$ in the flux tube model}

The flux tube model is extracted from the strong coupling limit of
the QCD lattice Hamiltonian\cite{isgur2,fluxtube}. In flux tube
mode, a meson consists of a quark and antiquark connected by
discretized quantum string. For conventional mesons, the string is
in its ground state. Vibrational excitation of the string
corresponds to the hybrid mesons\cite{isgur2,fluxtube}. The flux
tube model extends the $^3P_0$ model by including the dynamics of
the flux tube. This is done by including a factor that represents
the overlap of initial meson flux tube with those of the two final
mesons. Though the two models are not identical, their quantitative
futures are similar\cite{isgur2}, and the flux tube model coincides
with the $^3P_0$ model in the  limit of infinitely thick flux-tube.
In the rest frame of $A$, the decay amplitude of an initial meson A
into two final mesons $B$ and $C$ is,
\begin{eqnarray}
\nonumber{\cal M}(A\rightarrow B+C)&&=\int d^{3}\mathbf{r}_A\int
d^{3}\mathbf{y}\;\psi_A(\mathbf{r}_A)\;\exp(i\frac{M}{m+M}\mathbf{p}_B\cdot\mathbf{r}_A)\;
\gamma(\mathbf{r}_A,\mathbf{y})\\
\label{6}&&\times(i\nabla_{\mathbf{r}_B}+i\nabla_{\mathbf{r}_C}+\frac{2m\;\mathbf{p_B}}{m+M})
\psi^{*}_B(\mathbf{r}_B)\psi^{*}_C(\mathbf{r}_C)+(B\longleftrightarrow
C)
\end{eqnarray}
where both the flavor and spin overlaps have been omitted in the
above amplitude, and $\gamma(\mathbf{r}_A,\mathbf{y})$ is the
flux-tube overlap function, which measures the spatial dependence of
the pair creation amplitude. The initial quark and antiquark of the
initial meson $A$ are assumed to be of the same mass $M$, and $m$ is
the mass of the created quark pair. $\mathbf{y}$ is the pair
creation position, $\mathbf{r}_A$, $\mathbf{r}_B$ and $\mathbf{r}_C$
are respectively the quark-antiquark axes of A, B, and C mesons,
they are related by $\mathbf{r}_B=\mathbf{r}_A/2+\mathbf{y}$,
$\mathbf{r}_C=\mathbf{r}_A/2-\mathbf{y}$. For the conventional meson
decay, the flux-tube overlap function is usually chosen as the
following form\cite{isgur2},
\begin{equation}
\label{7}\gamma(\mathbf{r}_A,\mathbf{y})=A^{0}_{00}\;\sqrt{\frac{fb}{\pi}}\exp{(-\frac{fb}{2}\mathbf{y}^2_{\perp})}
\end{equation}
here
$\mathbf{y}_{\perp}=-(\mathbf{y}\times\mathbf{\hat{r}}_A)\times\mathbf{\hat{r}}_A$.
With these elements, the decay amplitude can be calculated
analytically under the simple harmonic oscillator wavefunction
approximation following the procedure for the calculation of widths
in Ref.\cite{isgur2}. The amplitudes for $2\;{^3D_1}\rightarrow
{^1S_0}+{^1S_0}$, $2\;{^3D_1}\rightarrow {^3S_1}+{^1S_0}$ etc are
derived by using  the harmonic oscillation parameter $\beta$ of the
outing mesons. The overall normalization factor $\gamma_0$ was
phenomenologically found to be equal to 0.64 for creating light
quark pairs\cite{isgur2,page,hybrid}. We take the string tension
$b=0.18$, and the constitute quark mass $m_u=m_d=0.33$GeV,
$m_s=0.55$GeV. As usual, the estimate value $f=1.1$ and
$A^{0}_{00}=1.0$ is used, a detailed discussion about these quantity
can be found in Ref.\cite{isgur2,paton}. The mesons masses are
chosen in the same way as in the above $^3P_0$ model case. The
numerical values for the partial decay width and total width are
shown in the third column of Table I, where we assume
$\beta_A=0.35$GeV.

We see that the overall behaviors of the decay modes in flux tube
model is similar to those in the $^3P_0$ model. $K(1460)K$,
$K_1(1270)K$, $K^{*}(1410)K$, $K^{*}K^{*}$, $KK$ are still the
dominant decay modes for $\beta_A=0.35$ GeV. Although there is large
difference comparing with the $^3P_0$ model in some channels, such
as $2{\;^3D_1}\rightarrow KK$, $2{\;^3D_1}\rightarrow K^{*}K$,
$2{\;^3D_1}\rightarrow K(1270)K$, which is due to the large strange
quark mass and the dynamical nodal suppression. This is strongly
supported by the fact that the partial decay width of
$2{\;^3D_1}\rightarrow\phi\eta$, $2{\;^3D_1}\rightarrow\phi\eta'$
and $2{\;^3D_1}\rightarrow h_1(1380)\eta$ are similar to those in
the $^3P_0$ model within $3\%$, even when considering variations in
$\beta_A$ and $M_A$, where a $s\bar{s}$ pair is created from the
vacuum. Additional evidence would be that the widths for the $S+S$
and $P+S$ final states($KK,K^{*}K,\phi\eta$ etc) in the flux tube
model would be the same as those in the $^3P_0$ model within $3\%$,
if we set $m_{u,d}=m_s$. For simplicity, the effect of
$m_{u,\;d}\neq m_s$ is ignored in the above $^3P_0$ model, following
Ref.\cite{barnes1,barnes2}, whereas this effect is included
explicitly in the flux tube model. A more delicate study of radial
strange quarkonium with the mass difference between the original
quark and the created quark considered may be necessary to improve
the overall agreement between the predictions and experimental data,
which has been done for the higher charmonium decay in the $^3P_0$
model\cite{barnes4}. For the interesting $K^{*}K^{*}$ mode,
$\Gamma_{F2}$ is predicted to the dominant as well, and the partial
widths ratios are the same as the $^3P_0$ model results.

In order to illustrate the parameter dependence of the model
prediction, we show the $\beta_A$ dependence of partial width and
total width for $2{\;^3D_1}$ $s\bar{s}$ decay respectively in Fig.5
and Fig.6, and the $M_A$ dependence is displayed in Fig.7 and Fig.8.
We can see that the $\beta_A$ and $M_A$ dependence is very similar
in the $^3P_0$ model and the flux tube model, we expect the
agreement will be improved further if the effect $m_{u,\;d}\neq m_s$
is considered. The width is predicted to be around $200$MeV, which
is large than the width of $1^{--}$ strangeonium hybrid\cite{ding}.
Although the predicted decay width is large than present
experimental value ($\Gamma=58\pm16\pm20 \rm{MeV}$), we cannot
exclude the $2{\;^3D_1}$ $s\bar{s}$ hypothesis considering the
uncertainties of the model and the experimental errors. The main
conclusions remain to be similar to the $^3P_0$ model's. Namely, the
modes $K(1460)K$, $K^{*}K^{*}$, $KK$ and $h_1(1380)\eta$ are crucial
in distinguishing between the $2{\;^3D_1}$ $s\bar{s}$ interpretation
and the strangeonium hybrid, which has a large branch ratio in the
former picture, but are forbidden instead in the latter .
\vspace{1cm}

\begin{figure}[hptb]
\centering
\begin{minipage}[t]{0.46\textwidth}
\centering
\includegraphics[width=8cm]{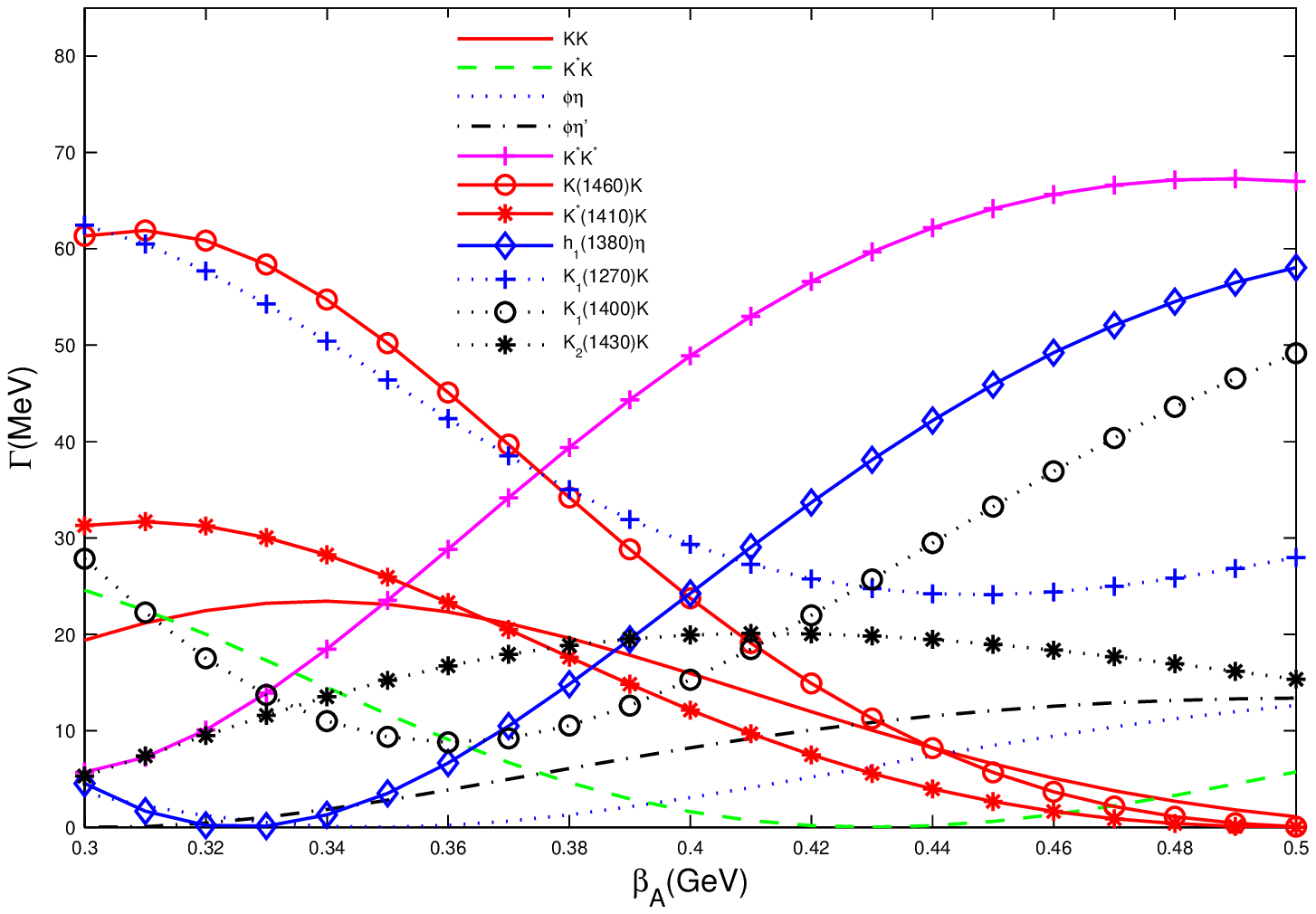}
\caption{The variation of Y(2175) partial decay widths with
$\beta_A$ as a $2{\;^3D_1}$ $s\bar{s}$ state in the flux tube
model.}
\end{minipage}%
\hspace{0.04\textwidth}%
\begin{minipage}[t]{0.46\textwidth}
\centering
\includegraphics[width=8cm]{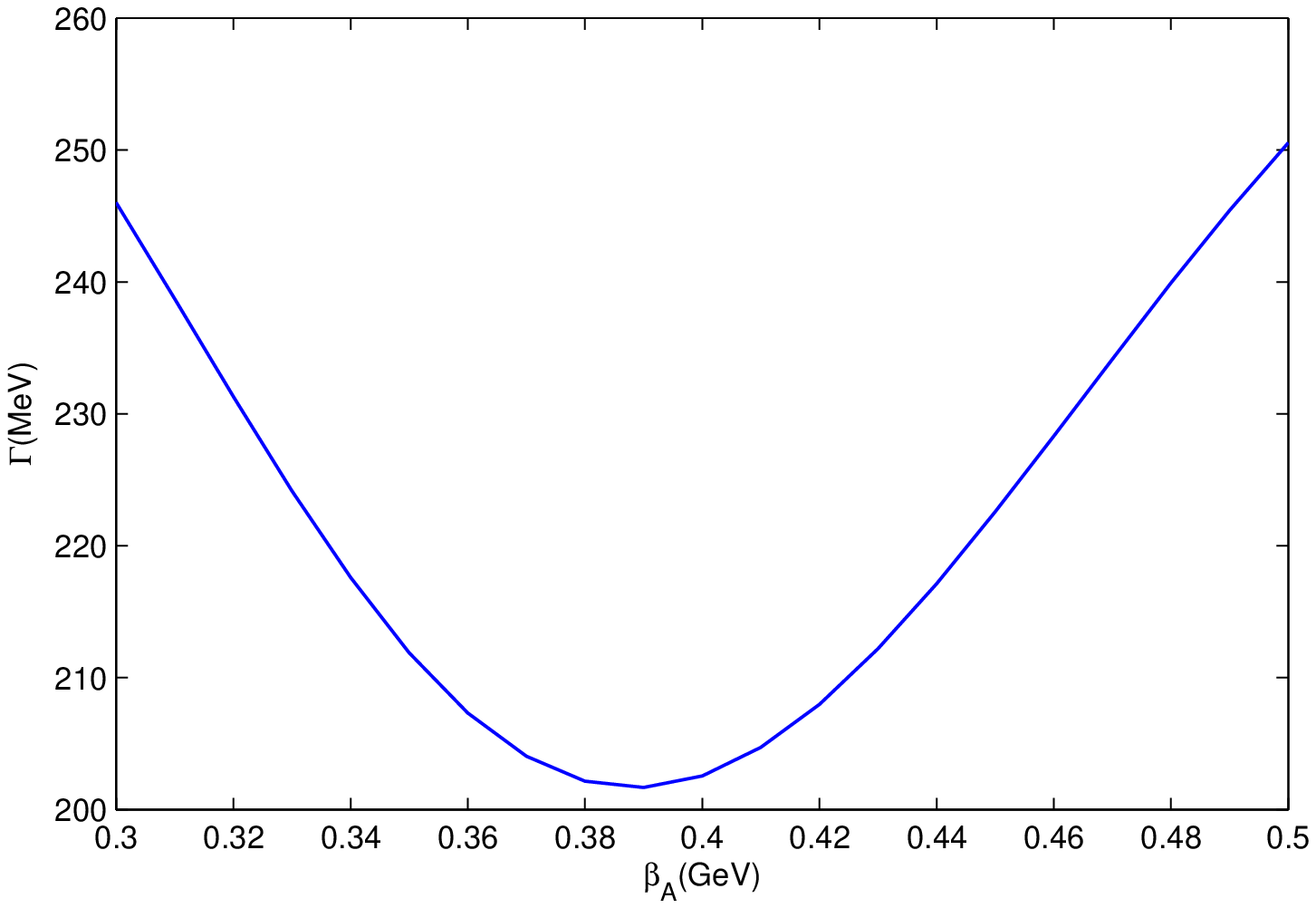}
\caption{Y(2175) total width dependence on $\beta_A$ as a
$2{\;^3D_1}$ $s\bar{s}$ state in the flux tube model.}
\end{minipage}
\end{figure}

\begin{figure}[hptb]
\centering
\begin{minipage}[t]{0.46\textwidth}
\centering
\includegraphics[width=8cm]{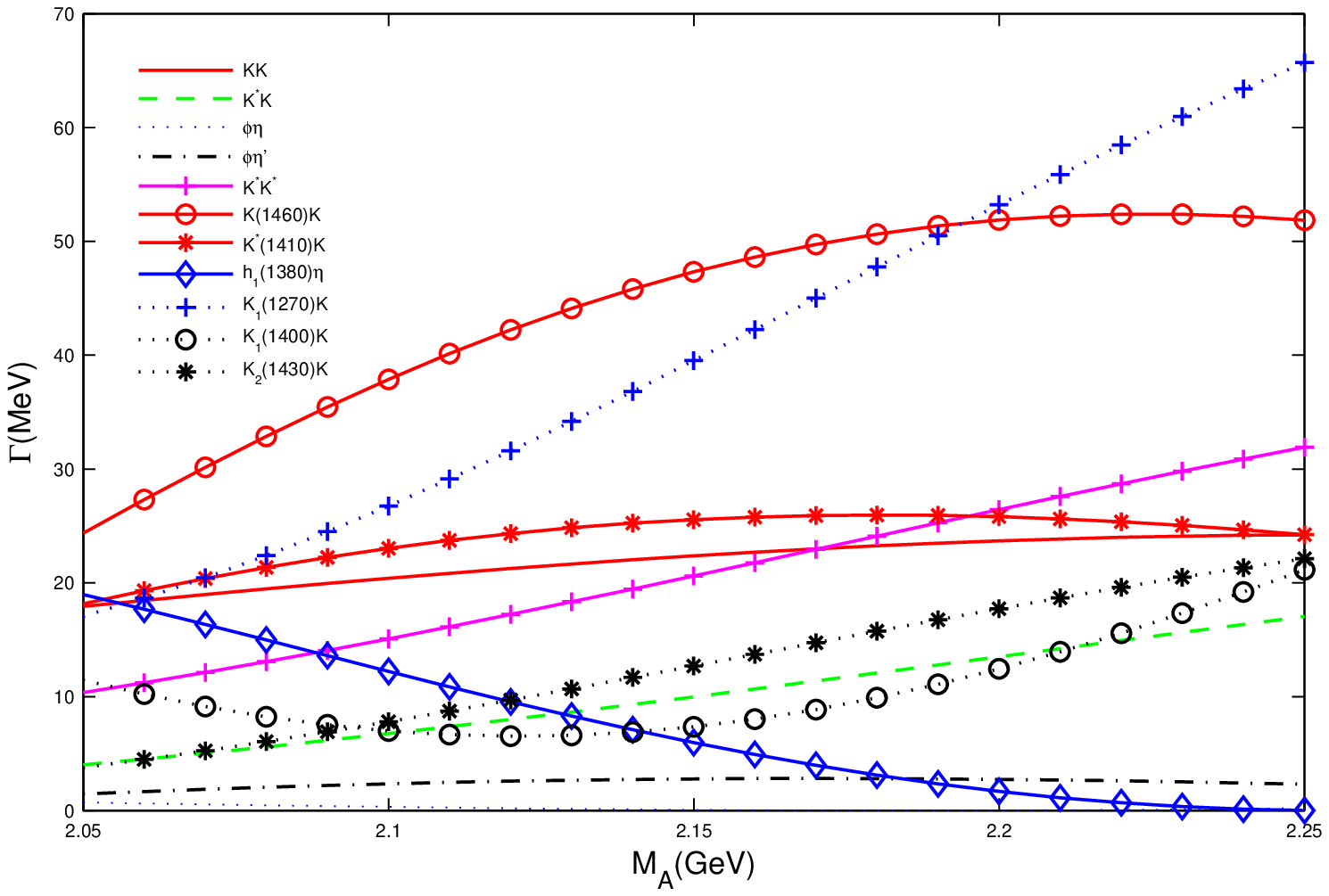}
\caption{Partial decay widths of Y(2175) as a $2{\;^3D_1}$
$s\bar{s}$ state at different initial state mass $M_A$ in the flux
tube model.}
\end{minipage}%
\hspace{0.04\textwidth}%
\begin{minipage}[t]{0.46\textwidth}
\centering
\includegraphics[width=8cm]{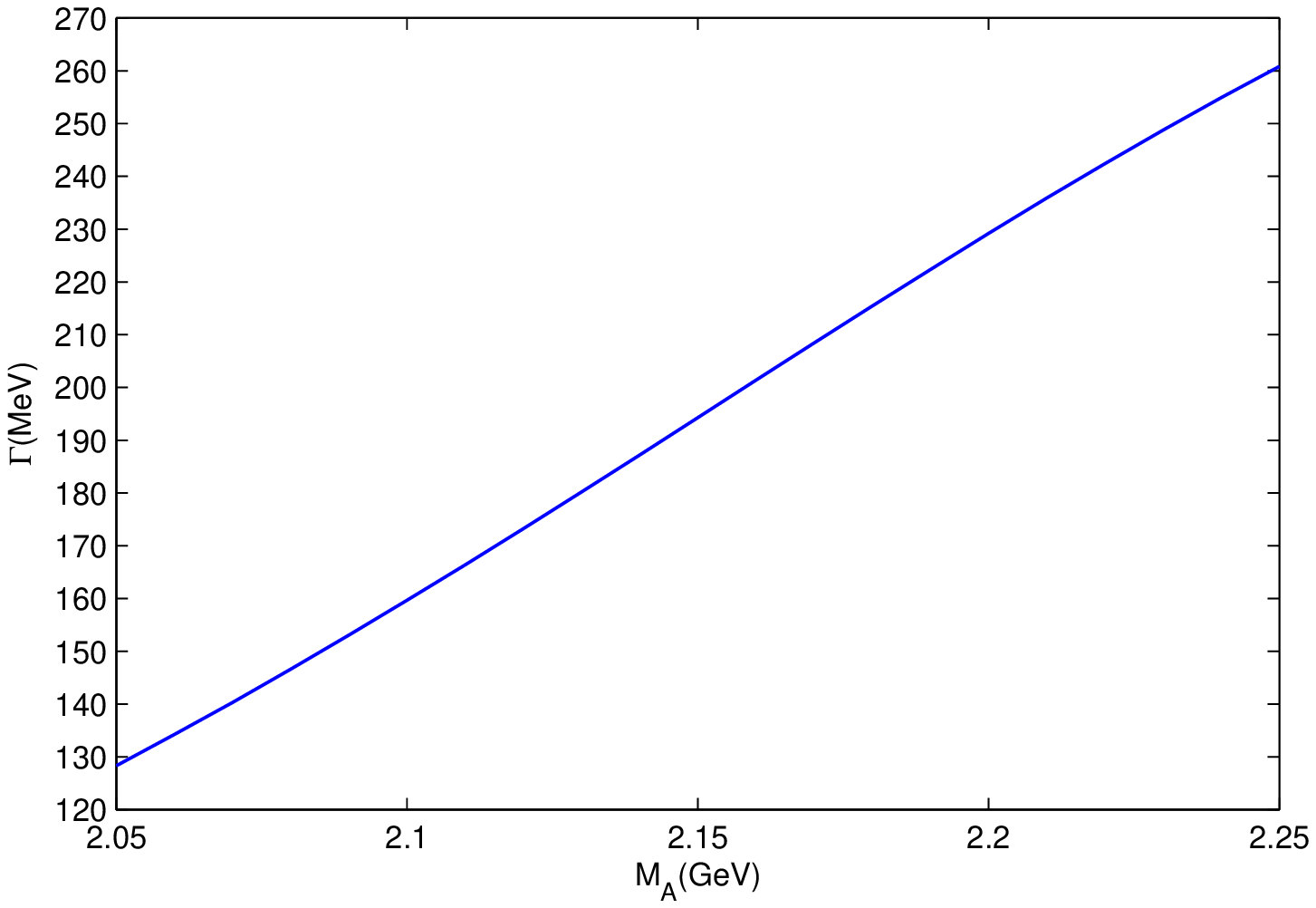}
\caption{Total widths of Y(2175) as a $2{\;^3D_1}$ $s\bar{s}$ state
at different initial state mass $M_A$ in the flux tube model.}
\end{minipage}
\end{figure}

\section{summary and discussions}

$Y(2175)$ has mass that is consistent with $2{\;^3D_1}$ $s\bar{s}$
and $3{\;^3S_1}$ $s\bar{s}$ meson. In this paper, we examine the
possibility that $Y(2175)$ is a $2{\;^3D_1}$ $s\bar{s}$ meson. We
would like to mention that the $3{\;^3S_1}$ $s\bar{s}$ state is
predicted to be a rather broad state by T.Barnes et al. (please see
the last column of Table I) so that it cannot be identified with
$Y(2175)$\cite{barnes2}. We have studied $2{\;^3D_1}$ $s\bar{s}$
decay from both the $^3P_0$ model and the flux tube model, and the
results are similar in the two models, despite the large difference
in some channels. This is due to large strange quark mass and
dynamic nodal suppression. The agreement will be better, if the mass
difference between the origin quarks in the initial state and the
created quarks is considered in the $^3P_0$ model.

It has been found that the difference between the decays of $1^{--}$
strangeonium hybrid and that of $2{\;^3D_1}$ $s\bar{s}$ is
significant. For the hybrid suggestion of \cite{ding},
 $Y(2175)\rightarrow K_1(1400)K,~
K_1(1270)K$ are the main decay channels, and the decay modes of
$Y(2175)\rightarrow KK,~ K^*K^*, K(1460)K, h_1(1380)\eta$ are
forbidden. For the $2 ^3D_1~ s\bar{s}$ scenario discussed in this
present paper, however, the decay modes of $Y(2175)\rightarrow KK,~
K^*K^*, K(1460)K, h_1(1380)\eta$ should be visible and the
corresponding decay widths are large in contrast to the hybrid
picture of $Y(2175)$. Therefore, we conclude that according to the
studies in \cite{ding} and the present paper the experimental search
of the $Y(2175)$'s decay modes $KK$, $K^{*}K^{*}$, $K(1460)K$
$h_1(1380)\eta$ is a criteria to identify the structure of
$Y(2175)$. In the other words, if these signals would be observed in
the experiment, $Y(2175)$ as a $2 ^3D_1~ s\bar{s}$ quarkonium is
favored, otherwise the interpretation of $Y(2175)$ as a hybrid is
preferred. At the present stage, because of the lack of such
experimental data, we cannot exclude the possibility that $Y(2175)$
is $2{\;^3D_1}$ $s\bar{s}$ meson. Obviously, it is crucial and
significant to detect the $Y(2175)$'s decay modes $KK$,
$K^{*}K^{*}$, $K(1460)K$, $h_1(1380)\eta$ experimentally in order to
identify whether $Y(2175)$ is a $2{\;^3D_1}$ $s\bar{s}$ quarkonium
or an exotic $s\bar{s}g$ hybrid meson.

The measurement of the ratios of the $K^{*}K^{*}$ partial decay
widths is a very important test of whether Y(2175) is a $2{\;^3D_1}$
or $3{\;^3S_1}$ $s\bar{s}$ state. Both $^3P_0$ model and the flux
tube model predict that $\Gamma_{F2}$ is the largest, and
$\Gamma_{P2}/\Gamma_{P0}=1/5$, provided that Y(2175) is a pure
$2{\;^3D_1}$ $s\bar{s}$ quarkonium. However, $\Gamma_{F2}$ is
predicted to be zero, and $\Gamma_{P2}/\Gamma_{P0}=20$ for the
$3{\;^3S_1}$ assignment.

\section *{ACKNOWLEDGEMENTS}
\indent We would like to acknowledge professor T. Barnes for his
valuable suggestions. This work is partially supported by National
Natural Science Foundation of China under Grant Numbers 90403021,
and by the PhD Program Funds 20020358040 of the Education Ministry
of China and KJCX2-SW-N10 of the Chinese Academy.

\end{document}